\documentclass[iop]{emulateapj} 
\usepackage{amssymb} 
\usepackage{times} 
\usepackage{epsfig} 
\usepackage{graphics}

\newcommand{\be}{\begin{equation}} 
\newcommand{\ee}{\end{equation}} 
\newcommand{\bea}{\begin{eqnarray}} 
\newcommand{\eea}{\end{eqnarray}} 
 
\slugcomment{ } 
 
\shorttitle{Magnetized NS atmospheres: beyond cold plasma approximation} 
\shortauthors{Suleimanov, Pavlov, \& Werner} 
 
\begin{document} 
 
\title{Magnetized neutron star atmospheres: beyond the cold plasma approximation 
    } 
 
 
\author{V. F. Suleimanov\altaffilmark{1,2}, G. G. Pavlov\altaffilmark{3,4}, and K. Werner\altaffilmark{1}} 
\affil{} 
\email{} 
 
 
\altaffiltext{1}{Institute for Astronomy and Astrophysics, Kepler Center for Astro and Particle Physics, 
   Eberhard Karls University, Sand 1, 72076 T\"ubingen, Germany; suleimanov@astro.uni-tuebingen.de} 
\altaffiltext{2}{Dept. of Astronomy, Kazan Federal University, Kremlevskaya 18, 420008 Kazan, Russia} 
\altaffiltext{3}{Pennsylvania State University, 525 Davey Lab., University Park, PA 16802; 
%, USA; 
pavlov@astro.psu.edu} 
\altaffiltext{4}{State Polytechnical University, Polytekhnicheskaya ul.\ 29, St.-Petersburg 195251, Russia} 
 
\begin{abstract} 
All the neutron star {  (NS) atmosphere} models published so far have been calculated 
in the  
``cold plasma approximation'', which 
neglects the relativistic effects in the radiative processes, such as  
cyclotron emission/absorption at harmonics of cyclotron frequency. 
Here we present new {  NS atmosphere} models which include such effects.  
We calculate a set of models  
 for effective temperatures $T_{\rm eff}=1$--3 MK 
and magnetic fields $B\sim 10^{10}$--$10^{11}$ G, typical for  
the so-called central compact objects (CCOs) in supernova 
remnants, for which the electron cyclotron energy $E_{c,e}$ and its first harmonics 
are in the observable soft X-ray range. Although the relativistic parameters, 
such as $kT_{\rm eff}/m_ec^2$ and $E_{c,e}/m_ec^2$, are very small for CCOs, the 
relativistic effects substantially change the emergent spectra at the 
cyclotron resonances, $E\approx sE_{c,e}$ ($s=1, 2,\ldots$). Although the 
cyclotron absorption features can form in a cold plasma due to the quantum 
oscillations of the free-free opacity, the shape and depth of these features change  
substantially if the relativistic effects are included. In particular, the features acquire deep 
Doppler cores, in which the angular distribution of the emergent intensity 
is quite different from that in the cold plasma approximation. The relative 
contributions of the Doppler cores to the equivalent widths of the features   
grow with increasing the quantization parameter 
$b_{\rm eff}\equiv E_{c,e}/kT_{\rm eff}$ and harmonic number $s$. 
The total equivalent widths of the features 
can reach $\sim$ 150--250 eV; 
they increase with growing $b_{\rm eff}$ and are smaller for higher harmonics. 
\end{abstract} 
 
 
\section{Introduction} 
 
Thanks to the high sensitivity and spectral resolution 
of the X-ray observatories {\sl Chandra} and {\sl XMM-Newton},  
 absorption features in thermal spectra of  
the so-called X-ray dim isolated neutron stars (XDINSs\footnote{These NSs are  
actually much brighter in X-rays than many other types  of isolated NSs, but we will call them XDINSs 
following a historical tradition.}) 
%or Magnificent Seven)  
and central compact objects (CCOs) in supernova remnants (SNRs) 
 have been discovered.  
The X-ray emission of these NSs is pulsed, and the measured period 
derivatives are consistent with strong (a few times $10^{13}$ G) 
magnetic fields in XDINSs (Haberl 2007; Kaplan 2008) and  
relatively weak ($\lesssim 10^{11}$ G) magnetic fields in CCOs 
(Halpern \& Gotthelf 2009). Therefore, the observed absorption features  
might be associated with ion cyclotron lines in XDINSs and electron cyclotron lines in CCOs.  
Correct identification of the observed features  can 
provide important information on the chemical composition, magnetic field and gravitational redshift at the NS surfaces. 
 
The most famous isolated NS with absorption features is 1E\,1207.4--5209 
(hereafter 1E\,1207),  
which is the CCO of the  
PKS 1209--51/52 SNR, 
  with properties similar to other CCOs (Pavlov et al.\ 2002, 2004; 
de Luca 2008; Halpern \& Gotthelf 2009). 
Two 
absorption features in the spectrum of this isolated NS, 
centered at about 0.7 and 1.4 keV,  
were discovered 
by Sanwal et al.\ (2002)  
in {\sl Chandra} observations, while later observations with {\sl XMM-Newton} suggested the presence of 
two more features, at 2.1 and 2.8 keV (Bignami et al.\ 2003).  
The period $P=0.424$ s  
was found by Zavlin et al.\ (2000) from X-ray pulsations. 
1E\,1207 shows a thermal-like 
X-ray spectrum with a black-body (BB) temperature $\approx$ 3 MK. 
The effective temperature $T_{\rm eff}$ may be substantially lower 
 ($\approx 1$ MK) if the spectrum is formed in a hydrogen atmosphere (Zavlin et al.\ 1998). 
 
It has been suggested that the harmonically spaced absorption features 
in 1E\,1207 are  
the electron cyclotron line and its harmonics (Bignami et al.\ 2003).  
At first glance, one should not expect 
 spectral features at  harmonics of the cyclotron 
energy  
because in a classical (non-quantum) plasma 
the harmonics are due to relativistic effects (e.g., the ratio of  
emissivities and opacities 
in consecutive harmonics is $\sim kT/m_ec^2$ at $kT\gtrsim E_{c,e}$), which are very small 
in the relatively cold CCO atmospheres. However, Suleimanov et al.\ 2010 (hereafter SPW10)  have demonstrated that,  
even in a cold plasma, 
harmonically spaced absorption lines  
can form due to quantum oscillations in the dependence of 
the free-free absorption on photon energy  
in a magnetic field (Pavlov \& Panov  1976; 
 {  see also a recent discussion of this process in Potekhin 2010}).  
The peak 
energies of these oscillations coincide with  
the cyclotron energy and its harmonics, 
 $E=s E_{c,e}=s \hbar eB/m_ec$ ($s=1, 2, \ldots$).  
For 1E\,1207, they correspond to 
the magnetic field  
$B_{\rm line} = 6\times 10^{10} (E_{c,e}^\infty/0.7\,{\rm keV}) (1+z)$ G  
($z$ is the gravitational redshift in the NS surface layers), 
consistent with   
$B_{\rm sd}=1\times 10^{11}$ G, which corresponds to 
 one of the two solutions obtained by Halpern  
\& Gotthelf (2011)  
for the pulsar's period 
derivative\footnote{This solution had been reported by Pavlov \& Luna 2009  
(see footnote \#15 of that work).}. 
 
The model atmospheres  
were calculated by SPW10 in  
the so-called cold plasma approximation (Ginzburg 1970), 
which implies, in particular, that effects associated with thermal motion 
of plasma particles  
can be neglected. All the previous magnetized NS atmosphere models, 
starting from the first work by Shibanov et al.\ (1992), 
have been calculated using this approximation.  
Without allowance for the above-mentioned quantum oscillations, there is 
only one resonance  
(the {  fundamental} cyclotron resonance at $E=E_{c,e}$) in the electron component of the opacity 
of a fully ionized ``cold'' plasma, 
and this resonance occurs only in one 
polarization mode, called the extraordinary mode (X-mode), while the energy 
dependence of the opacity of the ordinary polarization mode (O-mode) remains 
smooth  
(see Pavlov et al.\ 1980a; hereafter PMS80). 
The thermal motion  
leads to the cyclotron resonance 
($s=1$) in the 
O-mode (Pavlov et al.\ 1979), to emission/absorbtion of the cyclotron harmonics ($s>1$) in both 
polarization modes, and to the Doppler broadening of  
the cyclotron 
resonances (e.g.,  
PMS80); the contribution 
of these processes is determined by the ratio $kT/m_ec^2$. In addition to 
the thermal  
effects, quantum relativistic  
effects (whose contribution is  
determined by the ratio $E_{c,e}/m_ec^2$, 
 can be important when $E_{c,e} \gtrsim kT$ 
(Pavlov et al.\ 1980b; hereafter PSY80). 
  These effects  
have been ignored in the NS atmosphere modeling because the parameter 
$\xi_e \equiv {\rm max}(kT,E_{c,e})/m_ec^2$ is small ($\sim 10^{-3}$) 
 in atmospheres of isolated NSs. However, $\xi_e$ is not the only  
parameter that  
determines the  
strengths of the cyclotron resonances. For instance, 
the depth of the cyclotron absorption features depends on the ratio 
of the peak opacity to the opacity in the neighboring continuum,  
which may be $\gg 1$ even for a very small $\xi_e$. Therefore, 
to properly interpret the spectral features observed in CCOs, one 
should use NS atmosphere models calculated with account for the 
thermal motion of electrons and quantization of the electron motion 
in a magnetic field. 
 
 Here we present first calculations of  
realistic magnetized NS atmosphere models with these effects  
taken into account.  
We show that the fundamental resonance in  
the O-mode and at least two harmonics in both polarization modes are 
significant in the emergent spectra at the effective temperatures and magnetic fields typical for CCOs.  
In particular, the thermal motion effects lead to  
formations of relatively deep Doppler  
cores at the centers of  broad absorption features that arise 
due to quantum resonances in magnetic free-free absorption. 
These calculations demonstrate that the thermal motion and quantization  
must be taken into account 
for a quantitative description of the observed features in the spectra of CCOs. 
   
\section{Opacities in electron cyclotron line and harmonics} 
 
Our consideration of the thermal cyclotron opacities is based on  
the works of  
PMS80 and PSY80, 
where simple approximate  expressions for the cyclotron opacities were  
derived.  
 
 In PMS80,   
classic (non-quantum) relativistic opacities are considered with electron-ion 
collisions taken into account. 
{  According to that work, the shape of the cyclotron resonances can be 
described by the Voigt profile, which is the convolution of the 
Doppler and Lorentz profiles.} 
For $\Gamma_D \gg \Gamma_c + \Gamma_r$ 
(where $\Gamma_c$, $\Gamma_r$ and $\Gamma_D$ are the collisional, radiative, 
and Doppler widths of the cyclotron resonances), 
{  the Voigt  
profile consists of a Doppler core, independent of $\Gamma_c$ and 
$\Gamma_r$,  and Lorentz wings, which are virtually unaffected by the thermal 
motion. }  

In the atmospheres  
considered in our work, the quantum effects are significant because $E_{\rm c,e} \gtrsim kT_{\rm eff}$.  
{  The cyclotron absorption in the relativistic quantum case has been studied by 
PSY80 for a collisionless plasma. Since the free-free transitions, caused by electron-ion collisions, are  
not included in that work (as well as in later works on thermal cyclotron 
processes in quantizing magnetic fields), the results of PSY80 are applicable only 
in Doppler cores of cyclotron harmonics. However, similar to PMS80, 
one can use the cold 
plasma approximation outside the Doppler cores (i.e., in the Lorentz  
wings and the continuum), because the relativistic corrections are negligible there 
(at $kT_e \ll m_e c^2$). 
Moreover, since the thermal cyclotron opacity in the Doppler cores 
is much larger than the free-free opacity (except for the fundamental  
resonance in X-mode), and the thermal cyclotron opacity 
decreases exponentially in the wings of the resonances, one can calculate 
the total opacity as the sum of these two (see Figure 1 in PMS80).}    
Therefore, we use the following approach in our calculations.  
For each used photon energy $E$,  
we calculate separately the opacities in the cold plasma approximation,  
with the quantum oscillations in  
the magnetic free-free opacities taken into account  
(see details in Suleimanov et al. 2009 and SPW10), and the cyclotron opacities 
(with quantum effects taken into account)  
in the collisionless approximation (PSY80).  
{  Then,  
we simply add the collisionless opacities at all the harmonics (except 
for the fundamental resonance in X-mode) to the  
magnetic free-free opacities in the corresponding modes. For the fundamental  resonance in X-mode, we 
compare the two opacities  
and take the larger one as the actual opacity at this energy.  
The used approach only slightly distorts the 
frequency dependence of the  absorption coefficients in the transition regions 
between the cores and the wings, but this should not significantly affect 
the emergent spectrum because this transition occurs at different distances 
from the cyclotron resonances at different depths. 
} 
 
We perform all calculations  
  in the local thermodynamic equilibrium 
(LTE) approximation. 
{  This means that the cyclotron  
absorption is considered as a ``true absorption'', 
without scattering, and the cyclotron absorption does not contribute to the scattering part of 
the source function in the radiative transfer equations, which take into  
account the coherent electron scattering in the continuum only.} 
 
The approximate equations that we use to compute the cyclotron opacities are presented below (see PSY80 for details). 
They are not sufficiently accurate at angles $\theta$   
between the magnetic field and the photon wave vector close to 
$90^\circ$, i.e., at $|\cos \theta| \lesssim \beta$, where $\beta$ is the
ratio of the thermal velocity $v_T$ to the speed of light: 
\begin{equation} 
\beta = \frac{v_T}{c} = \left(\frac{2kT}{m_e c^2}\right)^{1/2} . 
\end{equation} 
As only a small amount of radiation is transferred at these angles,  
%therefore,  
we exclude  
%these limit angles  
them from our code and extrapolate the  
radiation field properties 
into this angle range. 
 
We calculate the opacities $k_j$ (cm$^2$ g$^{-1})$  
for polarization modes  
$j$=1 (X-mode) and $j=2$ (O-mode). 
The opacity $k_j$  
is proportional to the imaginary part, $\kappa_j$, of 
the complex refraction coefficient: 
% $\kappa_j$   
\be \label{u1} 
  k_j =  \frac{4\pi\nu}{c\rho} \kappa_j, 
\ee 
where $\nu$ is photon frequency and $\rho$ plasma density. We present equations for $\kappa_j$ below. 
 
\begin{figure} 
\centering 
\includegraphics[angle=0,scale=0.99]{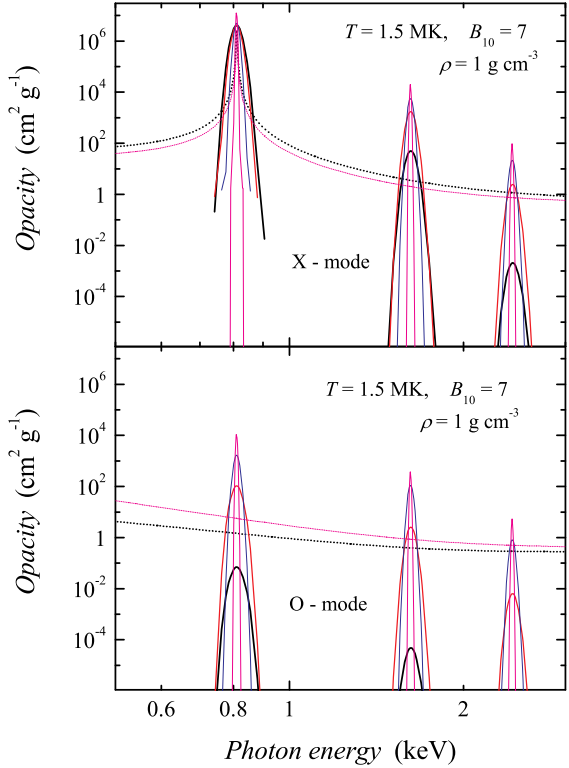} 
\caption{\label{fig1} 
Opacities in cyclotron lines and continuum opacities in  
two polarization modes (top and bottom panels) 
in a magnetized plasma with $T = 1.5$ MK, $\rho = 1$ g cm$^{-3}$,  
and $B = 7\times 10^{10}$ G, 
for  
four values of the angle $\theta$ between the magnetic field and the photon wave vector: 
5$^\circ$ (thickest black curves), 30$^\circ$ (thick red curves), 60$^\circ$   
(thin blue curves), and 80$^\circ$ (thinnest magenta curves). 
Continuum opacities for two values of $\theta$ are shown by dotted 
curves (thick black curves for  
$\theta$ = 5$^\circ$ and thin magenta curves for 80$^\circ$). 
} 
\end{figure}

An example of the computed continuum opacities, together with  
the opacities at the cyclotron line and its harmonics, 
% for  
%various $\theta$ are  
is shown in Figure \ref{fig1} for various $\theta$.  
%It is important  
Notice that, at some angles, the opacities at the fundamental resonance and  
 first two harmonics  
%at some angles  
are larger than the continuum opacities in both modes. This  
%means that we will obtain 
allows one to expect 
 significant absorption lines at these  
%photon energies  
resonances in the emergent spectra of magnetized NS model atmospheres.

\subsection{Opacities at the fundamental electron cyclotron resonance} 
 
According to PSY80, 
the equations for the normal mode opacities near the fundamental 
resonance are qualitatively different in the regimes of weak and strong vacuum  
polarization by magnetic field, i.e., at 
 small and large values of the parameter $V\beta\cos\theta\sin^2\theta$, where 
\be \label{2} 
  V = \frac{3\times 10^{28} {\rm cm^{-3}}}{n_e} \left(\frac{B}{B_Q}\right)^4, 
\ee 
$n_e$ is the electron number density, $B$ the magnetic field strength, 
and $B_Q = 4.414\times 10^{13}$ G is the magnetic field  
at which $E_{c,e}=m_ec^2$.  
In the case of CCOs, with their relatively weak magnetic fields, the regime of weak vacuum 
polarization is applicable throughout  the NS atmosphere. 
In this regime, the vacuum polarization does not affect the X-mode opacity: 
\be \label{u4} 
  \kappa_1 \approx   
%\frac{\omega_p^2}{\omega^2} 
\frac{\nu_p^2}{\nu^2} 
%(2\pi \nu)^2} 
\frac{1+\cos^2\theta}{4\beta |\cos\theta|} \sqrt{\pi}\, \exp(-x^2_1). 
\ee 
Here  
%$\omega=2\pi\nu$,  
$h\nu=E$ is the photon energy, and $\nu_p$ is the electron plasma frequency,  
\be \label{u5} 
   %\omega_p^2  
	\nu_p^2= \frac{e^2 n_e}{\pi m_e}\,. 
\ee 
%and $e$ the electron charge.  
The dimensionless energy shift $x_s$ relative to the central resonance energy in the $s$-th 
harmonic  
%with number $s$  
($s=1$ corresponds to the fundamental resonance) is defined by the following expression   
\be \label{u6} 
x_s = \frac{\xi -sb +  
%s|s| 
s^2\beta^2 b^2/4}{\varkappa} -  \frac{\varkappa}{4}, 
\ee 
where  
\bea \label{u7} 
b =  
\frac{h\nu_B}{kT},~~~h\nu_B=E_{c,e}, 
%\frac{E_{c,e}}{kT}, 
~~~\xi = \frac{h\nu}{kT}, \\ \nonumber 
~~~\varkappa = h\nu\left(\frac{2}{m_ec^2\,kT}\right)^{1/2} |\cos\theta|. 
\eea 
 
%The value of $\kappa_2$ for the O-mode is computed using this formula: 
In the same regime, the O-mode opacity is given by the following equation: 
\bea \label{u8} 
  \kappa_2 & = & \frac{\nu_p^2}{\nu^2} 
%(2\pi \nu)^2 
%\omega^2} 
\frac{\beta\sin^4\theta\sqrt{\pi}} {4(1+\cos^2\theta)|\cos\theta|} \left[\frac{b}{4}\tanh\frac{b}{2} \right. + \\ \nonumber 
& & \left. \frac{(3+\tan^2\theta-2V)^2} {\pi(1+\cos^2\theta)^2(H_1^2+F_1^2)} \cos^4\theta \right] \exp(-x_1^2). 
\eea 
Here $H_s\equiv H(x_s)$ and $F_s\equiv F(x_s)$ are the real and imaginary parts of the plasma dispersion function $W(x_s)$: 
\be \label{u9} 
     W(z)=H(z)+iF(z)= \frac{i}{\pi} \int_{-\infty}^{+\infty} \frac{e^{-u^2}}{z-u}\, du. 
\ee 
$H(z)$ is the well-known Voigt function,  proportional to the 
Doppler profile in our approximation,  
\be \label{u10} 
%       H(x_s+ia_s) = \frac{a_s}{\pi} \int_{-\infty}^{+\infty} \frac{e^{-y^2}}{(x_s-y)^2 - a_s^2}\, dy, 
H(x+iy) = \frac{y}{\pi} \int_{-\infty}^{+\infty} \frac{e^{-u^2}}{(x-u)^2 +
  y^2}\, du\,\, \stackrel{y \to 0}{\longrightarrow}\,\, e^{-x^2}, 
\ee 
while $F(z)$ is its Hilbert transform: 
\bea \label{u11} 
%       F(x_s+ia_s) = \frac{1}{\pi} \int_{-\infty}^{+\infty} \frac{e^{-y^2} (x_s-y)}{(x_s-y)^2 - a_s^2}\, dy. 
 F(x+iy) = \frac{1}{\pi} \int_{-\infty}^{+\infty} \frac{e^{-u^2} (x-u)}{(x-u)^2 + y^2}\, du\,\,  \\ \nonumber 
 \stackrel{y\to 0}{\longrightarrow}\,\,  
\frac{1}{\pi} \int_{-\infty}^{+\infty} \frac{e^{-u^2}}{x-u}\, du . 
\eea 

\subsection{Opacities in higher harmonics} 
 
%It is possible to compute values of  
The values of $\kappa_j$ in the harmonics $s>1$ 
are given by the following equation 
% for both modes using the same equation  
(signs ``plus'' and ``minus'' correspond to X-mode 
%, sign ``minus'' corresponds to  
and O-mode, respectively): 
\be \label{u15} 
 \kappa_j =  
%\frac{1}{8}  
\frac{\nu_p^2}{\nu^2}\, A \, (X \pm Y) \exp(-x_s^2), 
\ee 
where 
\be \label{u16} 
 X = 1+\cos^2\theta +  
%\xi 
\zeta\sin^2\theta, 
\ee 
\be \label{u17} 
 Y = (1+G^2)^{-1/2}\left[2|\cos\theta| + G(1-\zeta)\sin^2\theta)\right], 
\ee 
\be \label{u23} 
 G =  \frac{\sin^2\theta}{2|\cos\theta|} \frac{\nu_B^4 + V\nu^2(\nu^2-\nu_B^2)}{\nu_B^3\nu}, 
\ee 
\be \label{u18} 
 \zeta = \frac{\xi^2\beta^2\sin^2\theta}{4sb} \tanh\frac{b}{2}, 
\ee 
\bea \label{u19} 
 A =\frac{1}{8} \left(\frac{b}{1-e^{-b}}\right)^s \frac{1-e^{-\xi}}{\xi} \frac{s}{(s-1)!}  \times\\ \nonumber 
 \frac{\sqrt{\pi}}{\beta|\cos\theta|} \left(\frac{\beta\nu\sin\theta}{2\nu_B}\right)^{2s-2}. 
\eea

\section{Atmosphere models} 
 
We calculate the  
magnetic   
hydrogen atmosphere 
models  using our recently 
developed code 
(Suleimanov et al. 2009; SPW10), assuming the magnetic field normal to the stellar surface. 
We also assume that the atmosphere is fully ionized {  because, 
%and the magnetic field is normal to the  
%stellar surface. 
in the relatively low magnetic fields considered here, 
and temperatures of 
a few million Kelvin, the fraction of bound species in a hydrogen atmosphere 
is very low 
%, $\lesssim 0.1\%$  
(see, e.g., Pavlov et al.\ 1995).  
Moreover, all the hydrogen lines and 
photoionization jumps lie in far (or extreme) ultraviolet, unobservable 
because of the interstellar absorption (e.g., the energy of the strongest 
hydrogen line at $B=10^{11}$ G is about 26 eV, while the photoionization 
energy of the ground state is about 78 eV; see Roesner et al.\ 1984). 
Therefore, atomic lines are unobservable in CCOs, and bound species do 
not make a strong effect on the temperature run and the emergent spectrum.} 
% do not affect the 
%properties of the cyclotron harmonics we are interested in.} 
% {\bf The fully ionized plasma assumption is correct at the  
%relatively low magnetic fields considered here. The fraction of neutral hydrogen atoms is expected to be less  
%than 1\% (see Fig.\,2 in Potekhin \& Chabrier 2003).}   
%We also assume full ionization 
% and neglect vacuum 
%polarization (see, however, below) by the magnetic field,  
%because  relatively weak  
%magnetic fields are considered.  
 
\begin{figure} 
\centering 
\includegraphics[angle=0,scale=0.45]{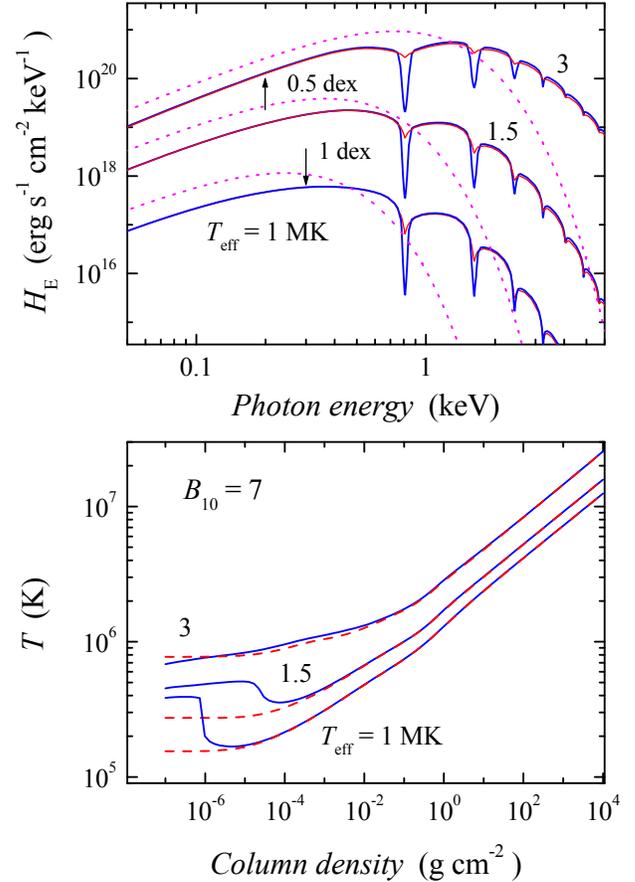} 
\caption{\label{fig2} 
{\it Top panel:} Emergent flux spectra for magnetic 
NS atmospheres  ($B=7\times 10^{10}$ G) in the cold plasma approximation 
(thin solid lines) and with  
%thermal  
the collisionless cyclotron opacities 
included  (thick solid curves)  
%and  
%without thermal cyclotron opacities (thin solid curves)  
for three 
effective temperatures, $T_{\rm eff}=1$, 1.5, and 3 MK. 
For clarity, the spectra for $T_{\rm eff}= 3$ and 1 MK are shifted 
along the ordinate axis by factors $10^{+0.5}$ and $10^{-1}$, respectively. 
The dotted curves show blackbody spectra for the same temperatures. 
{\it Bottom panel:} Temperature structures for the same models. 
Solid curves correspond to the models with  
%thermal  
collisionless 
cyclotron opacities included, dashed curves  to the models in the cold plasma approximation.  
Notice the overheating of the outer layers in the models with  
%thermal  
the {  collisionless} cyclotron 
opacities included.} 
%without thermal cyclotron opacities.} 
\end{figure}

We calculated two sets of atmosphere models with the same parameters as in SPW10. 
The surface gravitational acceleration $g = 10^{14}$ cm s$^{-2}$ for all the models.  
In the first set we fixed the 
magnetic field strength ($B = 7 \times 10^{10}$ G)  
and computed models with three different  
effective temperatures, $T_{\rm eff}=1$, 1.5, and 3 MK.  
%A comparison of e 
The emergent flux spectra and temperature structures for the models computed with and without 
%thermal  
the collisionless cyclotron opacities are shown in Figure \ref{fig2}. 
A common property 
of the emergent spectra for the models with the cyclotron  opacities 
are prominent 
absorption cores 
at the cyclotron energy 
and its harmonics, $E_s = sE_{c,e} = 0.81\,s$ keV ($s=1,2,\ldots$),  
in addition to the wide absorption wings 
caused by 
the quantum oscillations in the magnetic free-free opacities 
in the cold plasma approximation (SPW10). 
The equivalent width $w_s$  of the $s$-th absorption feature decreases with increasing $s$ 
as well as with increasing $T_{\rm eff}$ 
(i.e., with decreasing the effective 
quantization parameter, $b_{\rm eff}\equiv E_{c,e}/kT_{\rm eff} 
\approx 9.4$, 6.3, and 3.1 for the three models).  
%But  
However, these dependences are weaker than in  
%comparison to  
the  models computed in the cold plasma approximation. 
% without thermal cyclotron opacities. 
For instance,  
$w_1$= 243 (240), 213 (200), and 213 (180) eV at $T_{\rm eff}=1$, 1.5, and 3 MK, respectively, 
while $w_2$ = 234 (190), 205 (140) and 146 (55) eV at
 the same temperatures, where the numbers in parentheses are equivalent widths 
%. The corresponding $W_s$ in the 
%emergent spectra of  
for the models calculated in the cold plasma approximation. 
% without cyclotron opacities are given in parentheses.  

\begin{figure} 
\centering 
\includegraphics[angle=0,scale=0.45]{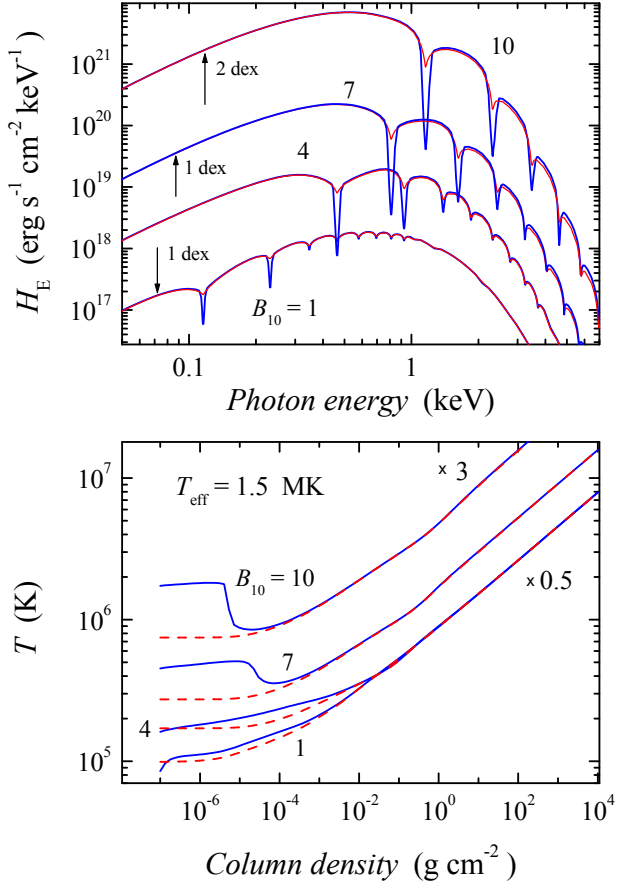} 
\caption{\label{fig3} 
{\it Top panel:} Emergent spectral fluxes  
for atmospheres with $T_{\rm eff}=1.5$ MK and 
 magnetic fields $B=1$, 4,  7 and 10 $\times 10^{10}$ G, 
calculated with and without 
including the  
collisionless cyclotron opacities (thick and and thin solid lines, 
respectively). 
For clarity, the spectra for $B_{10}= 10$, 7 and 1 are shifted 
along the ordinate axis by factors $10^{+2}$, $10^{+1}$ 
and $10^{-1}$, respectively.  
{\it Bottom panel:} 
Temperature structures for models with  
%(solid curves)  
and without  
%(dashed curves)  
%thermal  
the cyclotron opacities (solid and dashed curves, respectively), 
for the same parameters. For clarity, the temperature structures for 
$B_{10}= 10$, 4 and 1 are shifted 
along the ordinate axis by factors $3$, $0.5$ 
and $0.5$, respectively. 
} 
\end{figure}

The second set consists of four models with different magnetic field 
strengths ($B=1$, 4, 7, and 10 $\times 10^{10}$ G) at the same 
 effective temperature,  
$T_{\rm eff} = 1.5$ MK. 
The comparison of the emergent spectral fluxes for the models calculated with and without thermal cyclotron opacities  
is shown in  
the top panel of Figure \ref{fig3}.  
Again, the equivalent widths of the absorption features decrease with increasing harmonic number  
 and decreasing magnetic field strength. 
The difference of the emergent spectra computed with and without thermal cyclotron opacities 
is similar to that seen in Figure \ref{fig2}. 
In particular, the new models show substantially deeper cores of the 
absorption features at the cyclotron energy and its harmonics, while the continuum spectra remain virtually the same.

\begin{figure} 
\centering 
\includegraphics[angle=0,scale=0.45]{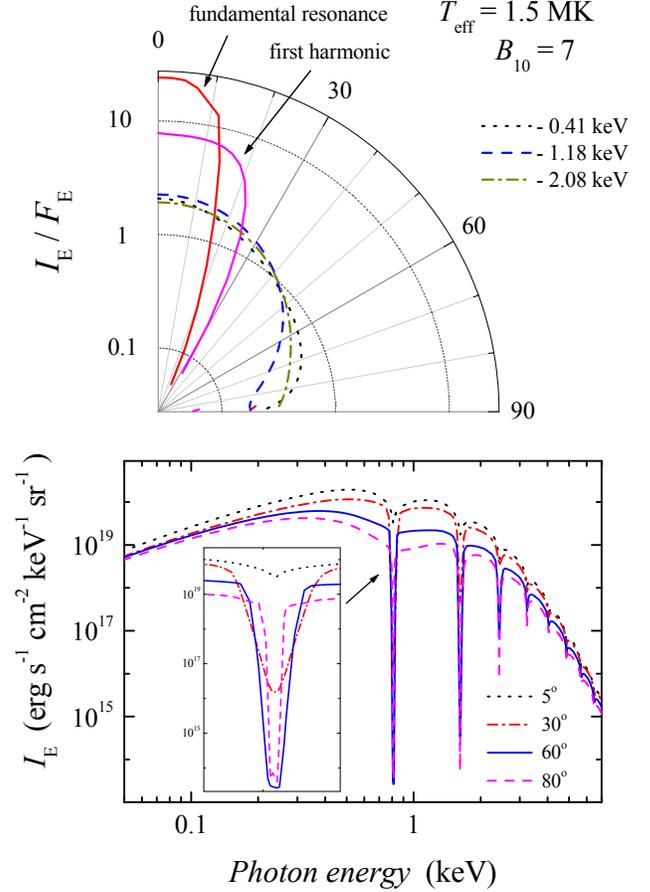} 
\caption{\label{fig4a} 
{\it Top panel:} Angular distribution of the emergent specific intensity at the  
fundamental resonance, first harmonic ($s=2$), and three continuum photon energies  
(indicated in the plot) for a NS atmosphere with  
$B=7\times 10^{10}$ G,  
$T_{\rm eff}=1.5$ MK.  
{\it Bottom panel:}  
Spectra of the emergent specific intensity  
for the same model at different 
angles to the surface normal (indicated in the plot).} 
\end{figure}

The bottom panel of Figure \ref{fig4a} shows 
spectra of emergent specific intensity  
for the model with  $T_{\rm eff} =$ 1.5 MK and  
$B = 7 \times 10^{10}$ G, for various angles $\theta$ 
between the line of sight and the normal to the surface. 
% normal 
%are shown in the bottom panel of Figure \ref{fig4a}. 
% 
The quantum oscillations together with deep cyclotron cores in the intensity spectra are seen at any 
angle. The cyclotron cores  
%vary significantly in comparison with  
%show a stronger dependence  
depend on $\theta$ stronger than the quantum oscillation wings.  
They are virtually invisible at small angles,  
in accordance with the disappearance of thermal cyclotron resonances 
at $\theta \to 0$ in the O-mode for $s=1$ and both modes for $s>1$ --- see Equations (8) and (17), 
%because the resonances in harmonics $s>1$ and  the opacity at the fundamental resonance  
%in O-mode (PSY80),  
and they become narrower and deeper with increasing $\theta$. 
The equivalent width of the  
%corresponding total  
absorption 
feature at the fundamental resonance depends on $\theta$ non-monotonously. 
For instance, 
%(e.g.,  
$w_1 \simeq$ 228 (220), 180 (160), 380 (320) and 400 (330) eV for $\theta=5^\circ$,  
$30^\circ$, $60^\circ$ and $80^\circ$, respectively 
%The corresponding $w_s$ in the intensity spectra of the models calculated without thermal 
%cyclotron opacities are again  
(the equivalent widths calculated in the cold plasma approximation 
are in parentheses). 
% (for this case the first and the last  
%values are computed at slightly different angles, 1$^\circ$ and 89$^\circ$ 
%[[DOES IT MAKE SENSE TO COMPARE $w_s$ CALCULATED FOR DIFFERENT 
%ANGLES?]]). 
 
In the top panel of the same Figure \ref{fig4a}, the angular distributions of the emergent 
intensity at five photon energies (near the fundamental resonance, the first harmonic, and 
% together with 
three continuum energies) are shown. Unlike the absorption features arising  
%due %to pure quantum oscillations 
in the cold plasma approximation (see SPW10),  the specific intensities at 
the cyclotron cores are  
%much more peaked towards 
stronger peaked towards the surface normal 
(i.e., towards the magnetic field direction) than those  
%  in comparison with those  
in the continuum.

%\begin{figure} 
%\centering 
%\includegraphics[angle=0,scale=0.6]{fig4b.ps} 
%\caption{\label{fig4b} 
%The same as in Fig.\,\ref{fig4a}, but the top plot 
%is in polar coordinates. 
%} 
%\end{figure} 
 
\begin{figure} 
\centering 
\includegraphics[angle=0,scale=0.45]{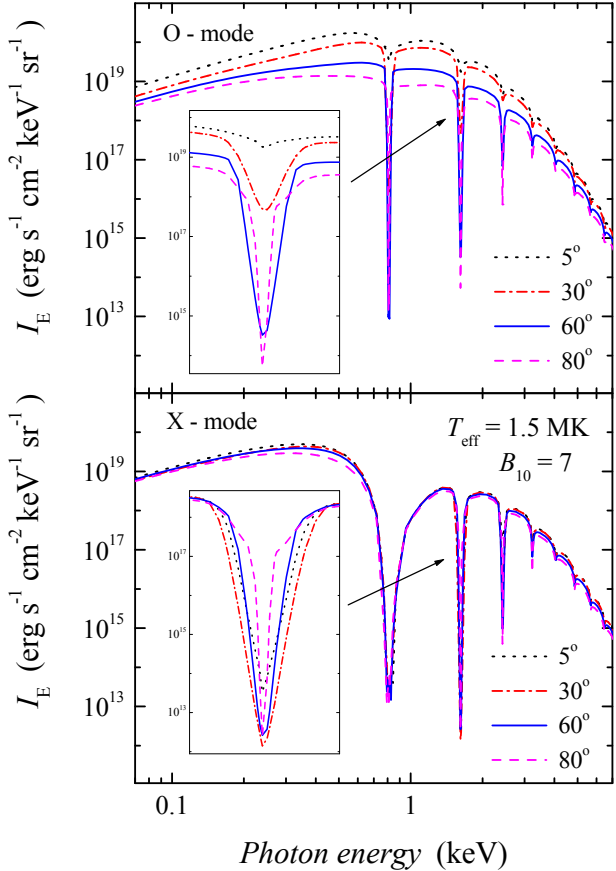} 
\caption{\label{fig4c} 
Spectra of the emergent specific intensity in the O-mode (top panel) 
and X-mode (bottom panel) 
for the same atmosphere model as in Figure \ref{fig4a}, at different 
angles to the surface normal (indicated in the plot).} 
\end{figure} 
 
Spectra of emergent specific intensity in the O- and X-modes,  
for the model with  the same parameters as in Figure \ref{fig4a}, 
%and for the same angles $\theta$ 
%between the line of sight and the surface normal 
are shown in Figure \ref{fig4c}. The intensity spectra in  
the X-mode depend on $\theta$ only slightly  
% are only slightly dependent on  
%$\theta$  
(bottom panel), while the  
%O-mode  
intensity  
spectra in the O-mode, which dominates at $E\gtrsim E_{c,e}$, 
vary with $\theta$ similar to 
% depend on  
%significantly decreases $\theta$  
%actually 
%identical  
%similar to the total spectra  
the intensity summed over polarizations (cf.\ Figure 4).

The bottom panels of Figures \ref{fig2} and \ref{fig3} show the temperature structures for 
the six atmosphere models in comparison with the temperature structures of the corresponding models  
computed in the cold plasma approximation. 
% without thermal cyclotron opacities.  
%The temperature structure  is 
%strongly affected by the thermal cyclotron opacities. 
% particularly 
%in the outer atosphere layers.  
%In particular, 
The main effect of the thermal cyclotron transitions is the temperature rise 
in outer atmospheric layers,  
%which is likely  
caused by an additional cyclotron heating due 
to the  
%cyclotron  
resonance in the O-mode opacity. 
%the thermal cyclotron resonance in the O-mode significantly raises 
%the temperature of upper atmospheric layers 
%at $ b_{\rm eff} \gtrsim 3$, 
%apparently because of cyclotron heating. 
%%, while outside of this range 
%the surface temperature becomes lower.  
%However, the effect of this overheating on the emergent radiation is not 
%significant. 
 
\begin{figure} 
\centering 
\includegraphics[angle=0,scale=0.99]{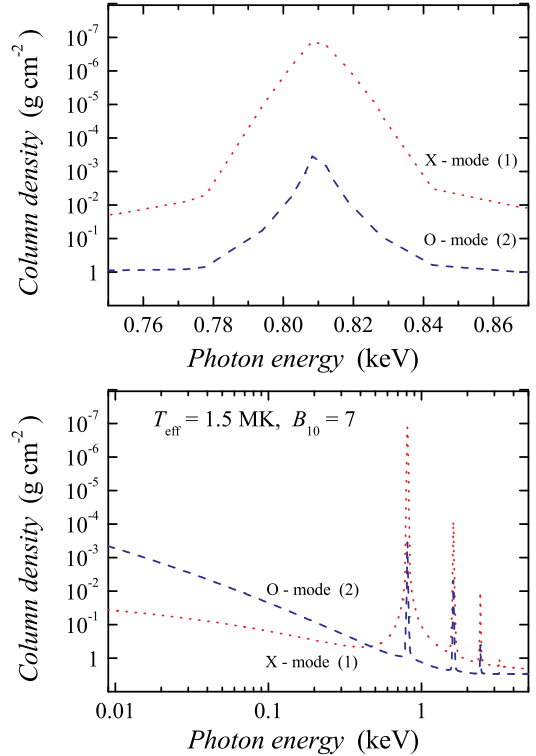} 
\caption{\label{fig8f} 
Photon energy versus flux formation depth (column density   
corresponding to $\tau^i_E$ =1) in the O- and X-modes for the  atmosphere 
with $T_{\rm eff}$ = 1.5 MK, $B_{\rm 10}$ = 7.  
In the top panel the energy band  
around the fundamental resonance  is shown in more detail.   
} 
\end{figure}

%In Figure \ref{fig8f} the  
The depths of the spectrum formation (the column densities which correspond to
$\tau^i_E=1$) 
for the new model are shown in Figure \ref{fig8f}. We see that in the region of cyclotron harmonics the 
O-mode comes from deeper (hotter) layers than the X-mode. Since these layers are 
much deeper than the outer layers  overheated by the cyclotron absorption (see 
Figures 2 and 3), 
%the modifi temperature profile is not manifested in the 
the overheating does not affect the emergent radiation summed over polarizations.

\begin{figure} 
\centering 
\includegraphics[angle=0,scale=0.45]{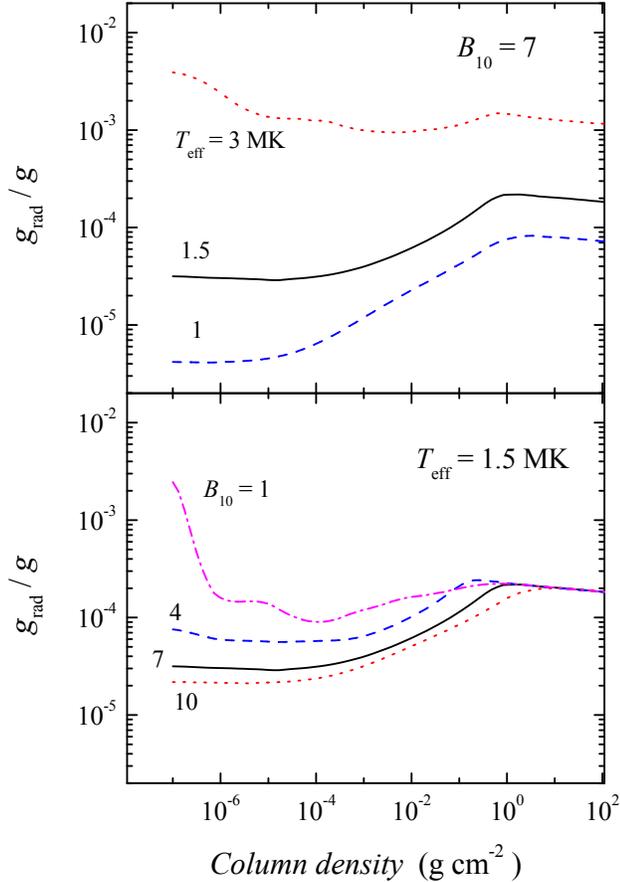} 
\caption{\label{fig6} 
Relative radiation accelerations in all the computed model atmospheres 
(see Figs.\,\ref{fig2} and \ref{fig3}). The atmosphere parameters  
are indicated in the panels. 
} 
\end{figure}

We also computed the radiative acceleration $g_{\rm rad}$, which is defined by 
the following equation 
% expression 
\be 
    g_{\rm rad} = \frac{2 \pi}{c} \sum_{i=1}^2 \int_0^{\infty}\, d\nu \, 
\int_{-1}^{+1} (k_{\nu}^i+\sigma_{\nu}^i) \, \mu \, I_{\nu}^i(\mu) \, d\mu, 
\ee 
where $\mu=\cos\theta$. The calculated $g_{\rm rad}$ values 
are smaller than 
% in comparison with the  
the surface gravity for all the computed model atmospheres (Figure \ref{fig6}) and can 
% may  
be ignored.   
 
\begin{figure} 
\centering 
\includegraphics[angle=0,scale=0.45]{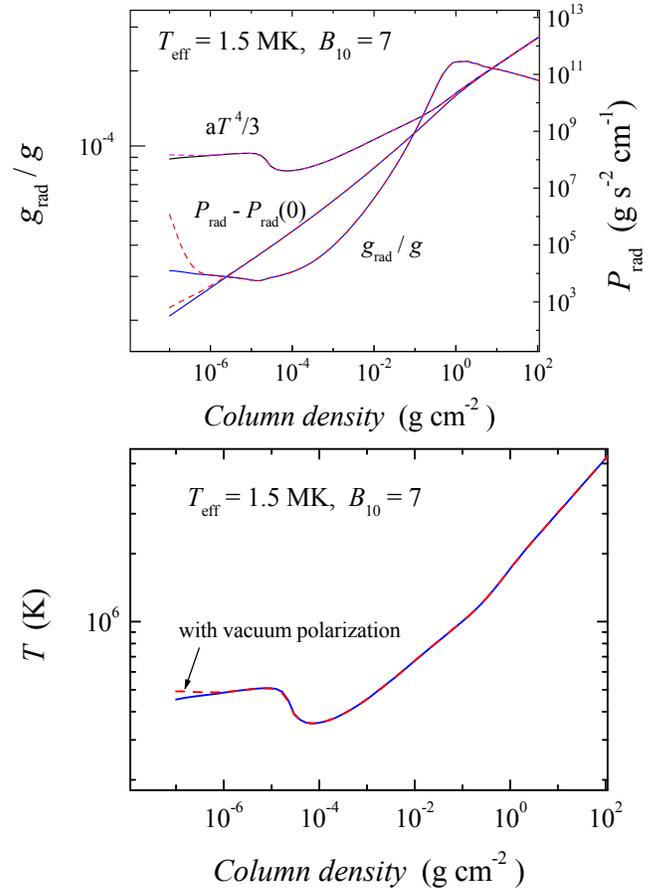} 
\caption{\label{fig5} 
Comparison of temperature structures (bottom panel) and relative  
accelerations due to radiation pressure (top panel) for models with (dashed curves)  
and without (solid curves) vacuum polarization taken into account. 
In the top panel, the corresponding radiation pressures are also shown. 
} 
\end{figure}

The vacuum polarization by the magnetic field might be significant  
in outer atmosphere layers,  
where the electron number density is very low.  
We have checked, however, that including the vacuum polarization into the  
%atmosphere structure  
calculations only slightly changes the temperature structure and the radiative
acceleration at 
the very surface (Figure \ref{fig5}), and  
it does not change the emergent spectrum at all. In the top panel of the same
figure, the radiation 
pressure distributions in the atmospheres are also shown, and the agreement
with the equilibrium thermodynamic value in the optically thick layers is demonstrated.   
 
\section{Discussion} 
 
We have presented the first 
computations of fully ionized hydrogen atmospheres of  
NSs for magnetic fields $B \sim 10^{10}$--$10^{11}$ G 
without using the cold plasma approximation. 
%, with thermal cyclotron harmonic opacities taken into account.  
For the magnetic fields chosen, 
 the electron cyclotron energy, $E_{c,e}\sim 0.1$--1 keV, 
is within the range of energies where the thermal emission from isolated 
NSs is usually observed. 
We have shown that  
%the deep Doppler cores, 
% in the first three harmonics, 
%caused by the thermal cyclotron opacities,  
%account for  
the thermal effects  
lead to formation of deep Doppler cores in the absorption  
features at the cyclotron energy and its harmonics, and to 
a sharper angular dependence of the emergent intensity at the cyclotron 
features.   
The main contribution to the equivalent widths of the cyclotron features still 
comes from the feature wings, which are due to the quantum oscillations in the
magnetic free-free opacity 
and can be considered in the cold plasma approximation. 
However, the  
contribution of the Doppler cores is significant, and it increases with  
increasing the quantization parameter 
$b_{\rm eff}$ and the harmonic number. 
The equivalent widths of the  
features 
reach $\sim$ 150--250 eV in the examples considered; they grow with increasing $b_{\rm eff}$ 
and are lower for higher harmonics. 
 
In our previous work (SPW10) we suggested that 
the harmonically spaced absorption features in the X-ray spectrum of the CCO  
1E\,1207  (Sanwal et al. 2002, Bignami et al. 2003) 
are due to the 
quantum oscillations in the magnetic free-free opacity. 
We ignored the relativistic effects, such as the  
thermal cyclotron processes, 
 because  the thermal and cyclotron energies are essentially nonrelativistic  
for   
 1E\,1207  
($kT_{\rm eff}/m_ec^2 < E_{c,e}/m_ec^2 \sim 10^{-3}$). However, 
 the detailed study presented here has demonstrated  
that these effects are important and must be included in the model atmosphere calculations. 
The reason is the large ratios  
of the cyclotron resonance opacities  
to the continuum opacity (e.g., $\sim 10^4$ and $\sim 10^1$ for the $s=2$  
and $s=3$ harmonics, respectively, for the atmosphere parameters  
relevant to the case of 1E\,1207). 
Therefore, the inclusion of the thermal cyclotron opacities in  
modeling atmospheres of 1E\,1207 and similar objects is needed. 
 
In future work we plan to compare the model spectra obtained by integrating the 
specific intensities over the visible NS surface  
(at different rotation 
phases and  for various orientations of the rotation and magnetic axes) 
 with the phase-dependent spectra of 1E\,1207 
to infer the  
properties of this NS. {   
%This  
The integration  
should lead to a broadening of cyclotron features 
% harmonics cores  
because of the nonuniformity of the surface magnetic field,  
%magnetic field distribution over NS surface,  
without dramatic changes of equivalent widths. We note that 
the pulsations detected in the radiation 
% only the 
%fluxes at absorption features in the  spectra  
of 1E\,1207  are strong only in the spectral features, 
%  phase-dependend,  
and we believe that the 
strong angular dependence of the modeled emergent intensity in the cores of
cyclotron harmonics (see Figures 4 and 5) 
can explain this fact and provide 
 a potentially powerful tool for the determination of the NS parameters.} 
 
In  
SPW10 we discussed the importance of quantum oscillations in the magnetic free-free opacity for the  
interpretation  of the X-ray spectra of some other CCOs (J1852+0040 in the
Kesteven 79 SNR and 
J0822--4300 in the Puppis A SNR) and 
 low-field radio pulsars. It is clear that the thermal cyclotron processes may also be significant for these objects 
and must be included in calculations of their model atmospheres. 
 
In the X-ray spectra of  
XDINSs, whose magnetic field are 2--3 orders of magnitude higher than in  
CCOs,  proton (or ion) cyclotron lines are probably observed.  
The thermal cyclotron processes  
should be less important for ions 
because the  
parameter $\xi_{p}\equiv {\rm max}(kT,E_{c,p})/m_pc^2$ is 
at least three orders of magnitude smaller 
than $\xi_e$, and the Doppler width of the ion cyclotron resonance  is $\approx 
40$ times smaller. 
However, the ratio of the ion resonance cyclotron opacity to the continuum 
opacity is also large for atmospheres of these NSs ($\gtrsim 10^7$, see, 
e.g. Figs.\,2 and 4 in Suleimanov et al.\ 2009). Therefore, a contribution 
of the ion cyclotron harmonics to the emergent XDINS model spectra cannot 
be excluded and have to be investigated\footnote{  Here we mean the thermal 
Doppler cores of the ion cyclotron harmonics, not the small quantum peaks in the magnetic free-free opacities 
%, which are insignificant  
(see Potekhin 2010 and references therein).}. 
 
In the model atmospheres computed here, the radiation transfer calculations 
were performed using the LTE approximation. Therefore, the source functions   
at the cyclotron line and its harmonics  were taken equal to  
the Planck function,  
and  
the photon scattering in the cyclotron resonances  
was  ignored.  
For a more accurate consideration, 
one should include non-coherent photon scattering in the cyclotron 
resonances. 
This is not simple (see, for example, Nagel 1981), and  
it has never been considered  
self-consistently.  
We plan to develop  
magnetized atmosphere models 
with non-coherent photon scattering in the cyclotron line and its harmonics  
in future work.

\acknowledgments 
VS thanks DFG for financial support (grant  
SFB/Transregio~7 ``Gravitational Wave Astronomy'').  
The work by GGP was partially supported by NASA (grant NNX09AC84G) 
and by the Ministry of Education and Science of the Russian Federation  
(contract 11.G34.31.0001).

%\clearpage 
 
\end{document}